\documentclass[fleqn,usenatbib]{mnras}

\usepackage{newtxtext,newtxmath}

\usepackage[T1]{fontenc}

\DeclareRobustCommand{\VAN}[3]{#2}
\let\VANthebibliography\thebibliography
\def\thebibliography{\DeclareRobustCommand{\VAN}[3]{##3}\VANthebibliography}

\makeatletter
\def\endfigure{\end@float}
\makeatother

\makeatletter
\def\endtable{\end@float}
\makeatother

\usepackage{graphicx}
\usepackage{amsmath}

\usepackage{amssymb}
\usepackage{multicol}       
\usepackage{booktabs}
\usepackage{newtxtext,newtxmath}
\usepackage{nicematrix}
\usepackage{caption}
\usepackage{enumitem}
\usepackage[utf8]{inputenc}
\usepackage{makecell}
\usepackage{threeparttable}
\usepackage{tabularray}
\usepackage{cleveref}
\usepackage{array}
\usepackage{float}
\usepackage{subfigure}
\usepackage{hyperref}
\usepackage{tabularx}
\usepackage{longtable}
\usepackage{lipsum}

\title[GC ULXs in NGC 1399]{Spectral Insights and Evolutionary Pathways of Globular Cluster ULX in NGC 1399: A Two-Decade X-ray and Optical Study}

\author[K. Oh et al.]{
Kwangmin Oh$^{1}$ \thanks{email:ohkwangm@msu.edu},
Kristen C. Dage$^{2,5}$\thanks{NASA Einstein Fellow},
Alexey Bobrick$^{3}$,
Elias Aydi$^{1,9}$\thanks{NASA Hubble Fellow}, 
Arash Bahramian$^{5}$,
Adelle J. Goodwin$^{5}$,
\newauthor
Daryl Haggard$^{6,7}$,
Jimmy Irwin$^{4}$,
Arunav Kundu$^{8}$,
Jay Strader$^{1}$,
Thomas J. Maccarone $^{9}$,
Stephen E. Zepf$^{1}$
\\
$^{1}$Michigan State University, Department of Physics and Astronomy, Michigan State University, East Lansing, MI 48824, USA\\
$^{2}$Wayne State University, Department of Physics \& Astronomy, Detroit, MI 48201, USA\\
$^{3}$Technion, Israel Institute of Technology, Physics Department, Haifa 32000, Israel\\
$^{4}$Department of Physics and Astronomy, University of Alabama, Box
870324, Tuscaloosa, Alabama, 35487, USA\\
$^{5}$ International Centre for Radio Astronomy Research $--$ Curtin University, GPO Box U1987, Perth, WA 6845, Australia\\
$^{6}$Department of Physics, McGill University, 3600 University Street, Montr\'eal, QC H3A 2T8, Canada\\
$^{7}$Trottier Space Institute at McGill 3550 University Street, Montr\'eal, QC H3A 2A7, Canada \\
$^{8}$Eureka Scientific, Inc., 2452 Delmer Street, Suite 100 Oakland, CA 94602, USA\\
$^{9}$Department of Physics, Box 41051, Science Building, Texas Tech University, Lubbock, TX 79409-1051, USA \\
}

\date{Accepted XXX. Received YYY; in original form ZZZ}

\pubyear{2024}

\begin{document}
\label{firstpage}
\pagerange{\pageref{firstpage}--\pageref{lastpage}}
\maketitle

\begin{abstract}

We present new multi-wavelength observations of two ultraluminous X-ray sources (ULXs) hosted by globular clusters (GCs) in the giant elliptical NGC 1399, focusing on CXO J0338318-352604 (GCU7), only the second GC ULX known to have luminous optical emission lines. Notably, only [N\textsc{ii}] and [O\textsc{iii}] emission is observed in the optical spectra, suggesting H-poor material. Previous work suggested the possibility that the properties of GCU7 could be explained by the tidal disruption of a horizontal branch star by an intermediate-mass black hole. We use new data to show that the lack of evolution in the X-ray or optical properties of the source over the last 20 years rules out this scenario. Instead, we use \textit{CLOUDY} simulations to demonstrate that the optical emission lines are consistent with an outflow from an ultra-compact X-ray binary where a compact object---likely a neutron star (NS)---is accreting above the Eddington limit from a helium white dwarf (He WD). This binary would have dynamically formed from a direct collision between a NS and a red giant, or else via an exchange interaction. The ULX is predicted to evolve to lower mass transfer rates over time and eventually become a doppelganger to the well-studied ultra-compact X-ray binaries in Galactic GCs such as 4U 1820--30. These results show the utility of using extragalactic GCs to study short-lived phases in dynamical binary evolution that occur too rarely to be observed in Galactic clusters.

\end{abstract}


\section{Introduction}

Ultraluminous X-ray sources (ULXs) are typically defined as X-ray point sources that are not located at the center of a galaxy and have an X-ray luminosity greater than the Eddington limit for stellar-mass black holes (BHs), e.g. $L_X$ $\gtrsim$ $10^{39}$ erg/s \citep{King_2023, Pinto_walton_2023}. ULXs are mainly detected in star-forming regions in younger galaxies \citep{kovlakas}, but starting with \cite{2007Natur.445..183M}, have been identified in extragalactic globular clusters (GCs) hosted by elliptical galaxies, out to as far as 70 Mpc \citep{2023MNRAS.518.3386T}. There are now around 20 ULXs known in extragalactic GCs \citep[][and references therein]{2023MNRAS.524.3662N}.

\cite{Dage19a} note that among these GC ULXs, two of them have been observed with optical emission above the GC continuum (originally discovered by \cite{zepf08} and \cite{Irwin}), while long-term X-ray observations of both show a soft X-ray spectrum for which the temperature does not change with changing $L_X$.   We note that a study by \cite{peacock_2012} searched for [OIII] emission unaffiliated with X-ray emission in GCs in NGC 4472, finding none, thus demonstrating that the chance precise alignment between optical and X-ray sources is highly unlikely. The remainder of the sample does not show optical emission and can have a more typical trend of increasing $kT$ with increasing $L_X$ \citep{2023MNRAS.518..855A}.

One of the two GC ULXs with observed optical emission is RZ2109, the first such source discovered by \cite{2007Natur.445..183M} in Virgo. A strong and broad ($\sim$1000 km/s) [O\textsc{iii}] 4959, 5007 \AA \hspace{0.01cm} emission line doublet was discovered by \cite{zepf08}, with no evidence of Balmer emission lines. The forbidden emission lines are indicative of strong outflows with accretion rates above Eddington, which may suggest that the primary is a stellar mass compact object.
\cite{2014ApJ...785..147S} set a stringent limit on the presence of hydrogen in this system, and thus the absence of hydrogen and the strong presence of oxygen strongly implicates a carbon-oxygen white dwarf (CO WD) as the donor star. 

RZ2109's X-ray luminosity peaks at around L$_X = 4 \times 10^{39}$ ergs/s at times, but is highly variable. Timing by \cite{2024MNRAS.529.1347D} suggests that this periodicity may be variable on the order of 1.3 days. They also show that the strength of the 5007 emission has been roughly constant over the 15 years it has been observed. 

The long time scales for observing both X-ray and optical emission from this system strongly rule out alternatives such as a nova illuminated by an X-ray binary \citep{2012MNRAS.423.1144R} or a tidal disruption event (TDE) \citep{2011ApJ...726...34C}. Instead, both X-ray and optical properties are very consistent with a young ultracompact X-ray binary (UCXB) with a stellar-mass BH primary and a CO WD secondary \citep{2024MNRAS.529.1347D}. Specifically, the mass of the CO WD secondary requires that the primary must be a stellar-mass BH in order to form a stable accreting system \citep{2017MNRAS.467.3556B,2017ApJ...851L...4C}. 

UCXBs have previously been suggested as the counterparts of bright X-ray sources in GCs in elliptical galaxies \citep{2004ApJ...607L.119B}. Additionally, within our Galaxy, UCXBs are significantly overproduced in GCs compared to the field: Galactic GCs contain almost one-third of persistent UCXBs \citep{2023A&A...677A.186A} while having less than $10^{-3}$ of the Galactic stellar mass, e.g., \citet{2018MNRAS.478.1520B}.

The other GC ULX with observed optical emission lines, and the main subject of this study, CXO J033831.8-352604, is located in a GC associated with the elliptical galaxy NGC~1399 \citep{Irwin}. Both [N \textsc{ii}] and [O\textsc{iii}] have been observed, with line luminosity on the order of $10^{36}$erg/s and FWHM of 140 km/s (instrumental FWHM = 55 km/s). Recent work by \cite{Zhou_2023} identified helium emission lines in NGC 247 ULX-1, a field ULX in a spiral galaxy, attributed to the outer accretion disk or wind of a helium star donor with a FWHM of about 200 km/s. Although field ULXs form in different environments than GC ULXs, this highlights the range of optical signatures observed in ultracompact systems.
    
It was initially suggested that GCU7 could be explained by an intermediate-mass black hole (IMBH) tidally disrupting a WD, although the cause of the [N\textsc{ii}] emission remained a mystery \citep{Irwin}. \citet{2011MNRAS.410L..32M} suggested that the system may involve a compact object binary ULX photoionizing a nearby R Cor Bor star within the same GC, though not directly interacting with it. \cite{2012MNRAS.424.1268C} then proposed the partial tidal disruption of a red clump horizontal branch star by a lower mass BH. Here, we investigate whether there is a different model of this source that is consistent with the lack of long-term evolution of the X-ray and optical emission reported here and also in which the ULX and the material producing the forbidden optical emission lines arise from the same binary system.

Given the strong observational similarities between CXO J033831.8-352604 and RZ2109, as described above, and RZ2109's strong status as a UCXB candidate, CXO J033831.8-352604 may also be explained through the lens of UCXB theory. Another compelling comparison point is \cite{2006MNRAS.370..255N}, whose optical spectroscopy finds that N is abundant in UCXB spectra. A recent study by \cite{Barra_2024} also identified nitrogen-rich material in the X-ray emitting plasma of Holmberg II X-1, highlighting the prevalence of nitrogen enrichment in accreting systems, albeit in a different environment and donor context. In this paper, we present new X-ray and optical observations of CXO J033831.8-352604, and discuss it in the context of UCXB evolutionary theory. 

To better understand the evolutionary environment of UCXBs, we conducted a preliminary analysis using Monte Carlo simulations, following a recent study by \cite{Oh_2024}, which found that the dynamical status of GCs affects their cataclysmic variable (CV) populations. Dense GCs tend to have more bright CVs in the X-ray due to the frequent dynamical interactions of compact objects in their core regions. The study classified simulated clusters based on their core density evolutionary tracks. Clusters with expanding cores and decreasing core density were designated as Class I. Clusters with increasing core density, which indicates core collapse and promotes stellar interactions and binary formation, were classified as Class II. Finally, Class III clusters, which have undergone core collapse, showed either stabilized or slightly decreasing core density during re-expansion, reflecting the most dynamically evolved cluster. The study compared simulated results with observations and found consistent classifications. Thus, we adopted these classes and simulated GC data to understand the UCXB production environment.

In Section \ref{sec:data}, we provide new X-ray and optical observations from \textit{Chandra}, \textit{Gemini}, and \textit{SOAR}. In Section \ref{sec:discussion}, we show the results of modeling the optical emission to determine if a He WD or He star could produce the observed emission. A second GC ULX, CXO 033832.6-352705 is also present in the NGC 1399 observations. We briefly discuss the source in light of a new X-ray and an archival optical spectrum from 2006. We also further discuss these results in light of UCXB theory, and whether such a system is likely to have formed as part of the GC's rich evolutionary process. In Section \ref{sec:summary}, we discuss the implications of these results. 

\section{Data and Analysis}
\label{sec:data}

Two GC ULXs,  CXO J033831.8-352604 (which we will refer to as GCU7) and CXO 033832.6-352705 (GCU8) were observed by \textit{Chandra} 13 times between 1999 and 2023 (\autoref{tab:chandra-observations}). Observed properties of the sources are listed in \autoref{table:properties}. Optical spectra of GCU7 have been obtained eight times with three observatories from 2008 until 2023 (\autoref{table:eqwid}). GCU8 was observed by the Magellan II Clay telescope on Nov 26th 2006. 

\subsection{X-ray Observation Log}

\begin{minipage}{0.95\columnwidth}
\setlength{\tabcolsep}{5pt}
\renewcommand{\arraystretch}{1.2}
\begin{tabularx}{0.95\columnwidth}{XXXc}
\hline
\textbf{ObsID} & \textbf{Exposure} & \textbf{Counts/s} & \textbf{Date} \\
\hline
320 & 3.38 & 152.79   & 1999-10-18 \\
319 & 56.04 & 11.47   & 2000-01-18 \\
239 & 3.6 & 7.27      & 2000-01-19 \\
49898 & 13.08 & 9.8   & 2000-06-15 \\
240 & 43.53 & 10.07   & 2000-06-16 \\
2389 & 14.67 & 10.69  & 2001-05-08 \\
4172 & 44.5 & 4.96    & 2003-05-26 \\
9530 & 59.35 & 15.79  & 2008-06-08 \\
14527 & 27.79 & 11.06 & 2013-07-01 \\
16639 & 29.67 & 11.03 & 2014-10-12 \\
14529 & 31.62 & 10.98 & 2015-11-06 \\
27748 & 19.71 & 9.43  & 2023-07-31 \\
26675 & 20.38 & 8.68  & 2023-08-02 \\
\hline
\end{tabularx}
\captionof{table}{Summary of Chandra X-ray Observations: observational ID (ObsID), total exposure time in kiloseconds (ks), the average count rate in counts per second, and the date of observation.}
\label{tab:chandra-observations}
\end{minipage}

\begin{table*}
\begin{tabular}{lllllllll}
\hline \hline
Name & RA         & Dec       & Peak Lx      &  Spectral Shape & g (extinction corrected)     & z   (extinction corrected)   & g-z  & $r_h$      \\ \hline
GCU7 & 03:38:31.8 & -35:26:04 & $2.3\times 10^{39}$ erg/s & kT $\sim$ 0.4 keV                    & 22.645 & 21.275 & 1.37 & 2.11 pc \\
GCU8 & 03:38:32.6 & -35:27:05 & $5\times 10^{39}$ erg/s   & $\Gamma \sim$ 1.2                    & 22.091 & 20.491 & 1.6  & 1.94 pc \\\hline
\end{tabular}
\caption{GC ULXs in NGC 1399 and their observed properties. We omit  CXOKMZJ033831.7-353058 \citep{Shih} from this study as it has not been detected in X-ray since 2003, and/or it has been outside the field of view of many of the recent \textit{Chandra} observations. RA and Dec are given in J2000 coordinates. Peak L$_X$ represents the maximum observed X-ray luminosity. The spectral shape shows the best-fit model. g, z shows the extinction-corrected magnitudes, and g–z is given for the colour index. r$_h$ represents a half-light radius in pc. Optical properties are adopted from \citet{jordan_2015}.}
\label{table:properties}
\end{table*}

\begin{table}

\centering
\begin{tabular}{lccc}
\hline \hline
\textbf{Instrument} & \textbf{Date} & \textbf{[O\textsc{iii}]} & \textbf{[N\textsc{ii}]} \\ 
\hline
Magellan & Oct 27, 2008 &  71.2 $\pm$ 9.0   & 50.1 $\pm$ 5.7      \\ 
Gemini   & Sep  1, 2011 &  66.2 $\pm$ 10.5  & -                   \\ 
SOAR     & Aug 27, 2019 &  78.6 $\pm$ 27.7  & 47.4 $\pm$ 6.7      \\ 
SOAR     & Oct 15, 2020 &  64.2 $\pm$ 19.5  & 48.5 $\pm$ 2.8      \\ 
SOAR     & Dec 12, 2020 &  90.9 $\pm$ 27.0  & 53.8 $\pm$ 7.7      \\ 
Gemini   & Dec 30, 2021 &  61.6 $\pm$ 7.7   & 45.2 $\pm$ 3.8      \\ 
SOAR     & Dec 25, 2022 &  -                & 43.7 $\pm$ 5.8      \\ 
SOAR     & Oct 14, 2023 &  65.4 $\pm$ 30.0  & 54.0 $\pm$ 10.2     \\ 
\hline
\end{tabular}
\caption{Observation details and equivalent widths of the [O\textsc{iii}] and [N\textsc{ii}] emission lines for GCU7. The 2011 Gemini observation did not extend above 6000 Angstrom.}
\label{table:eqwid}
\end{table}

\subsection{X-ray observations}
The data for NGC 1399 were collected using the Advanced CCD Imaging Spectrometer (ACIS) as shown in \autoref{tab:chandra-observations}. We processed these data using the script \texttt{chandra\_repro} provided by the \textit{Chandra} Interactive Analysis of Observations (CIAO) software version 4.15.1, applying the calibration database (CALDB) version 4.10.7 released on 2023-09-14. For each GC image, we employed the wavelet detection algorithm (a CIAO tool, \texttt{wavdetect}) using wavelet scales of 2, 4, 8, 16, 24, 32, and 48 pixels. These values allow us to detect both compact and more extended X-ray features across a range of sizes. We set a significance threshold of $10^{-6}$, which equates to approximately one false alarm per execution of the algorithm on an image of $1k \times 1k$ pixels. We employed the CIAO \texttt{srcflux} script to compute the unabsorbed net X-ray fluxes, denoted as $f_x$, within the energy range of 0.3 to 8 keV by using an absorbed power-law model. We fixed the line-of-sight absorption, $n_H$, to a value of $1.14 \times 10^{20}$ cm$^{-2}$, obtained from \textsc{colden}\footnote{\url{https://cxc.harvard.edu/toolkit/colden.jsp}}. 

\begin{figure}
\includegraphics[width=\columnwidth]{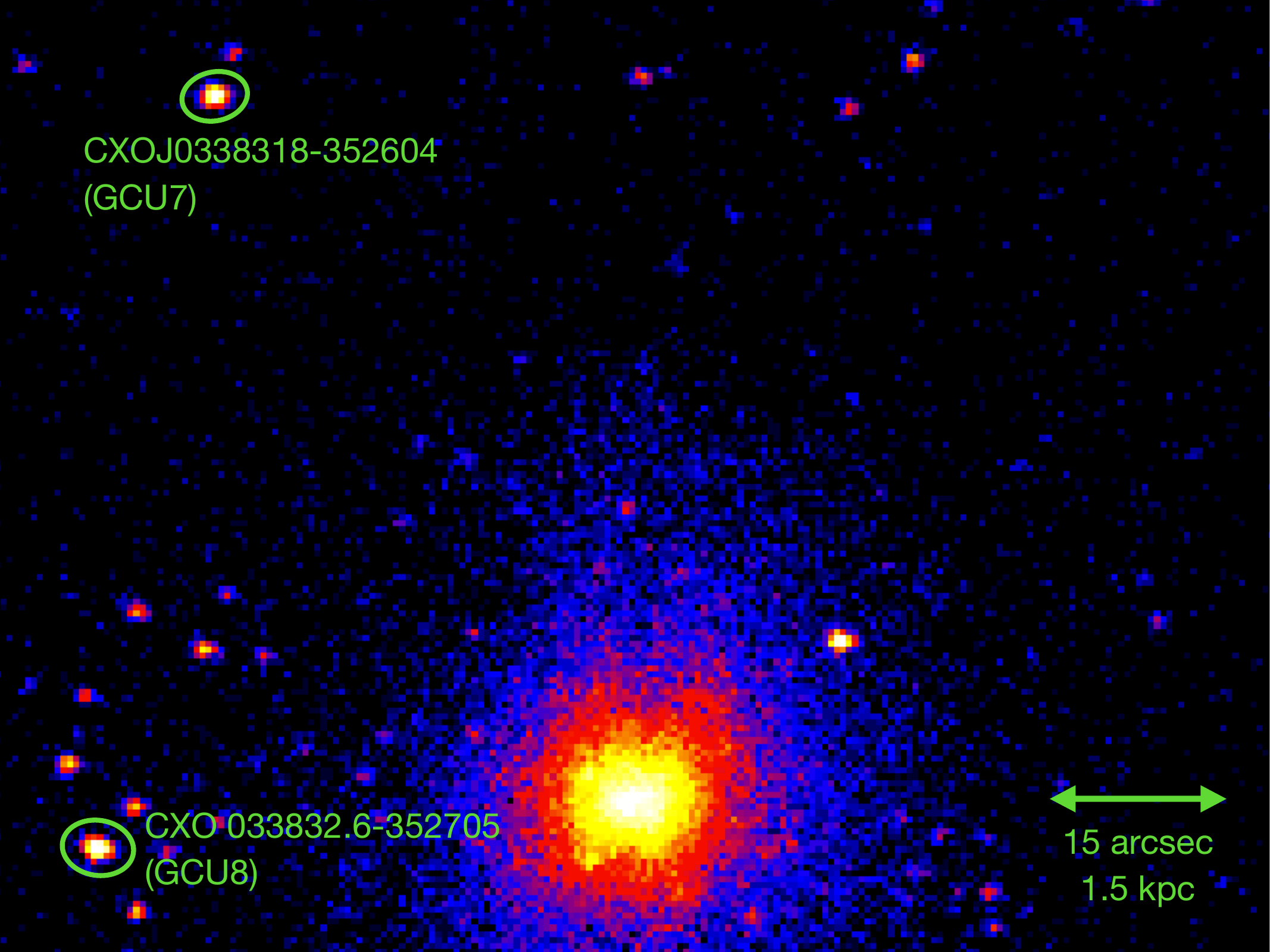}
\caption{A stacked X-ray image combining 13 \textit{Chandra} X-ray observations, processed in the energy range of 0.5–8.0 keV. The ULXs analyzed in this study, GCU7 and GCU8, are highlighted by the green regions.}
\label{fig:chandra_image}
\end{figure}

To enhance the image quality and maximize the signal-to-noise ratio, \autoref{fig:chandra_image} shows a stacked X-ray image, combining 13 individual \textit{Chandra} observations in the 0.5 to 8.0 keV energy range. This approach maximizes resolution and provides a clear view of the ULX sources, GCU7 and GCU8, analyzed in this study.

\subsection{X-ray Spectral Analysis}
We examined the spectrum of the most recent observations of NGC 1399, OBSIDs 26675 and 27748. To analyze the spectral features in the X-ray observations, we used the CIAO tool {\tt specextract} to extract the spectra and calculate the response files. For the background spectra, we sampled from nearby source-free regions close to the corresponding X-ray sources. All spectral fittings were performed using XSPEC version 12.13.1. 

We analyzed the spectra of each source separately in each observation. By fitting the source spectrum with disk blackbody and power-law models ($N_{H}$ fixed at $1.14\times10^{20}$~cm$^{-2}$), we found the spectral parameters inferred from both observations to be consistent within the tolerance of their $1\sigma$ statistical uncertainties. These models were chosen since they effectively capture the thermal emission from the accretion disk and the non-thermal processes. Given the lack of spectral variability, we proceeded to fit the spectra from \textit{ObsID} 26675 and \textit{ObsID} 27748 simultaneously to optimize the photon statistics, using the best-fit models from \cite{Dage19a}.

\begin{figure}
\includegraphics[width=\columnwidth]{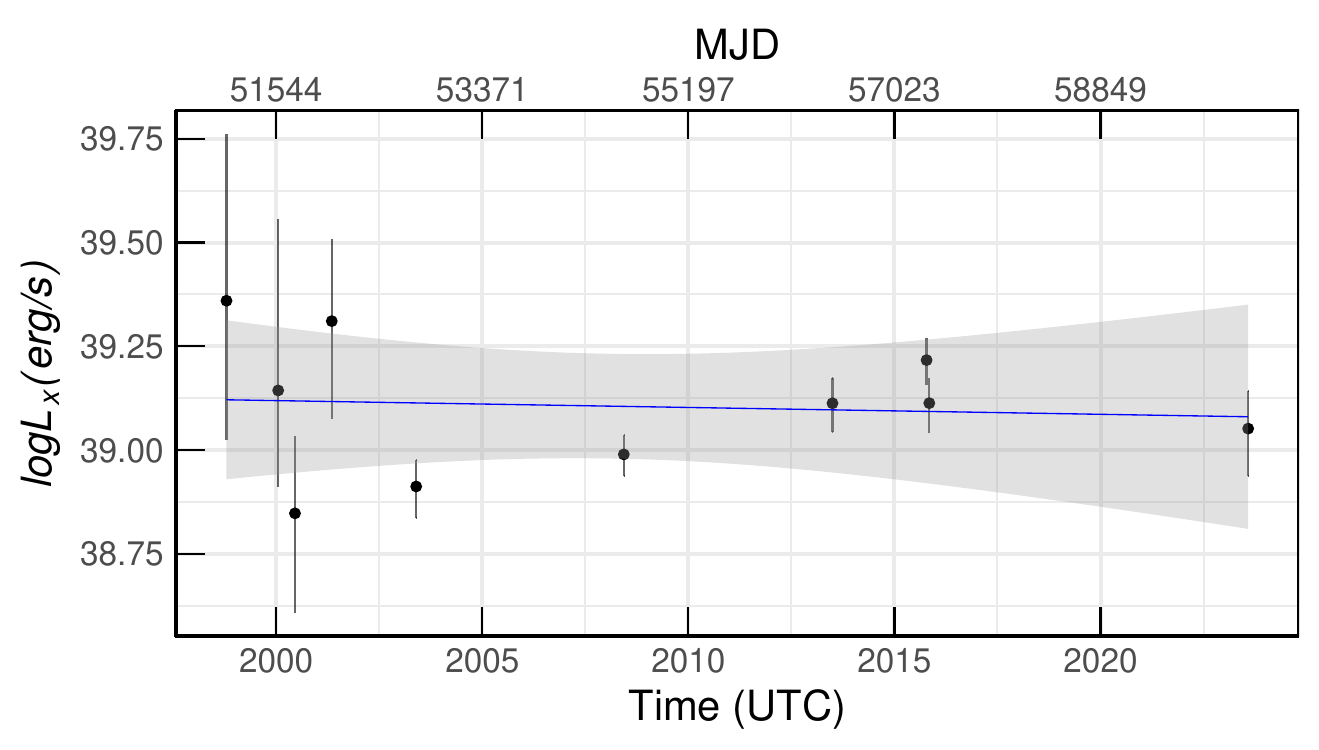}
\caption{Luminosity evolution profile of the ULX source GCU7 over a 20-year of \textit{Chandra} X-ray observations. The X-ray luminosity, calculated from the unabsorbed flux in the 0.5–8.0 keV energy range, remains relatively stable over a period of two decades. The shaded region indicates the 95\% confidence interval of the regression line.}
\label{fig:variability}
\end{figure}

\autoref{fig:variability} presents the long-term variability of GCU7 over a 20-year period. The luminosity was calculated from unabsorbed X-ray fluxes in the 0.5–8.0 keV energy range. As shown in the figure, GCU7 exhibits relatively stable X-ray luminosity with no significant fluctuations over the two decades of observations. To simplify the analysis, closely spaced observations were combined.

\begin{table*}
\centering
\renewcommand{\arraystretch}{1.3}
 \begin{tabular}{c|cccc|cccc}
    \hline
    \hline
    & \multicolumn{4} {c|}{Disk blackbody} & \multicolumn{4} {c}{Power-law}\\
    \hline
Source & {$T_{in}$} & Disk norm & {$\chi^{2}$/$\nu$} & Disk flux & {$\Gamma$} & {Norm} & {$\chi^{2}$/$\nu$} & PL flux \\
& (keV) 
& ($10^{-3}$) 
&  
& ($10^{-14}$ ergs cm$^{-2}$s$^{-1}$)
&
& ($10^{-5}$ $keV^{-1}$cm$^{-2}$$s^{-1}$)  
&
& ($10^{-14}$ ergs cm$^{-2}$s$^{-1}$)\\
\hline
GCU7 & 0.57 $\pm {0.09}$   & 9.56 $\pm {7.20}$ &  15.63/14 & 1.61 $^{+0.09}_{-0.81}$ &
 3.12 $\pm {0.43}$   & 0.94 $\pm {0.25}$    & 13.91/14 & 2.10 $^{+0.63}_{-0.57}$ \\
GCU8 & 1.53 $\pm {0.21}$   & 0.63 $\pm {0.32}$ &  18.00/14 & 6.53 $^{+0.07}_{-2.23}$ &
 1.58 $\pm {0.17}$   & 1.26 $\pm {0.21}$    & 17.37/14 & 7.79 $^{+0.64}_{-0.61}$ \\
\hline
\hline
  \end{tabular}
  \caption{Spectral fit results for the ULX sources GCU7 and GCU8 in NGC 1399. The table shows the inner disk temperature ($T_{in}$), disk normalization, and flux for the disk blackbody model, along with the photon index ($\Gamma$), normalization, and flux for the power-law model. The $\chi^{2}/\nu$ values indicate the goodness-of-fit for both models.}
  \label{tab:myfits}
\end{table*}

\subsection{Optical Observations and Spectral Analysis}
\begin{figure}
\centering
\subfigure{
\includegraphics[width=0.95\columnwidth]{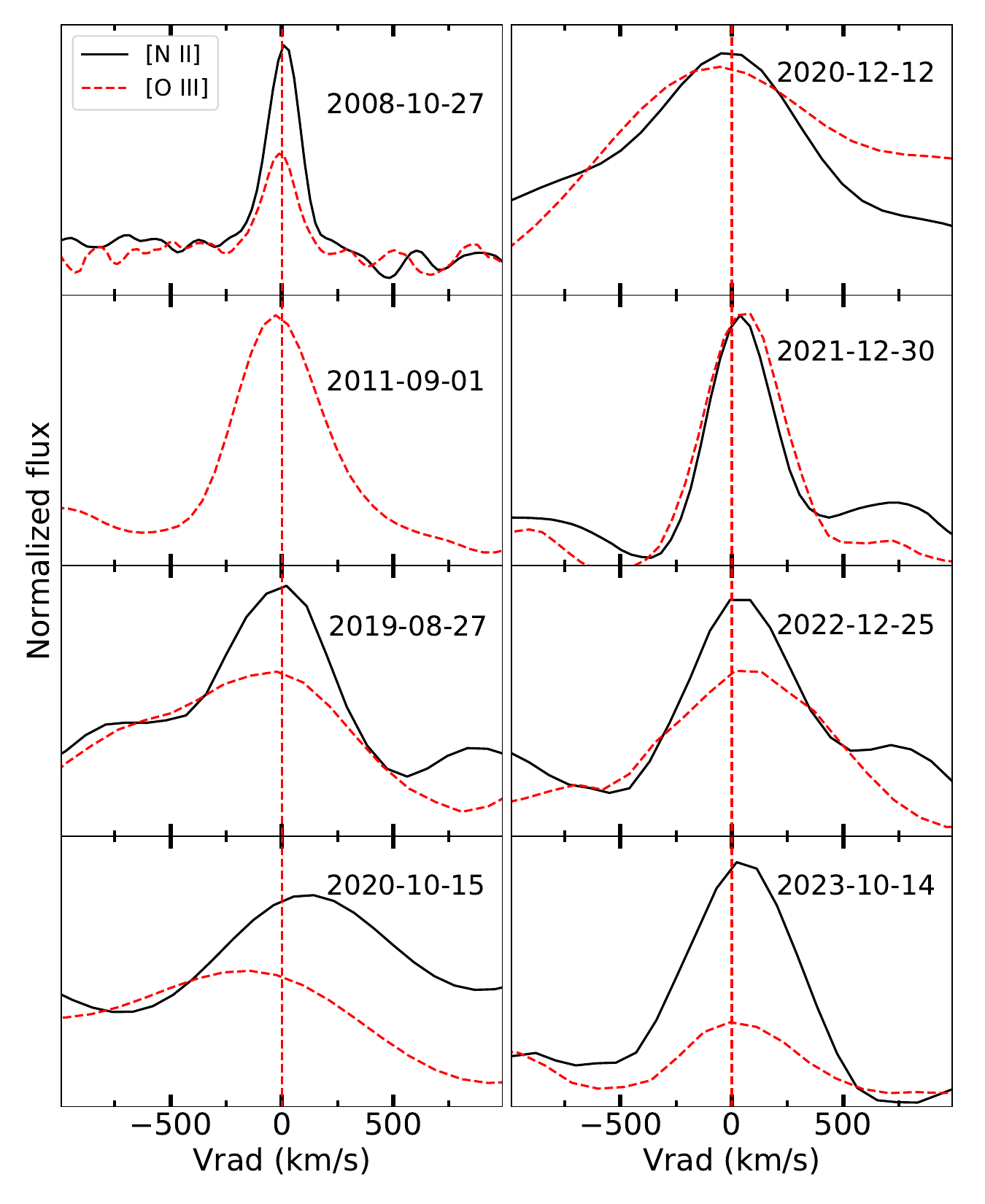}
}
\subfigure{
\includegraphics[width=0.95\columnwidth]{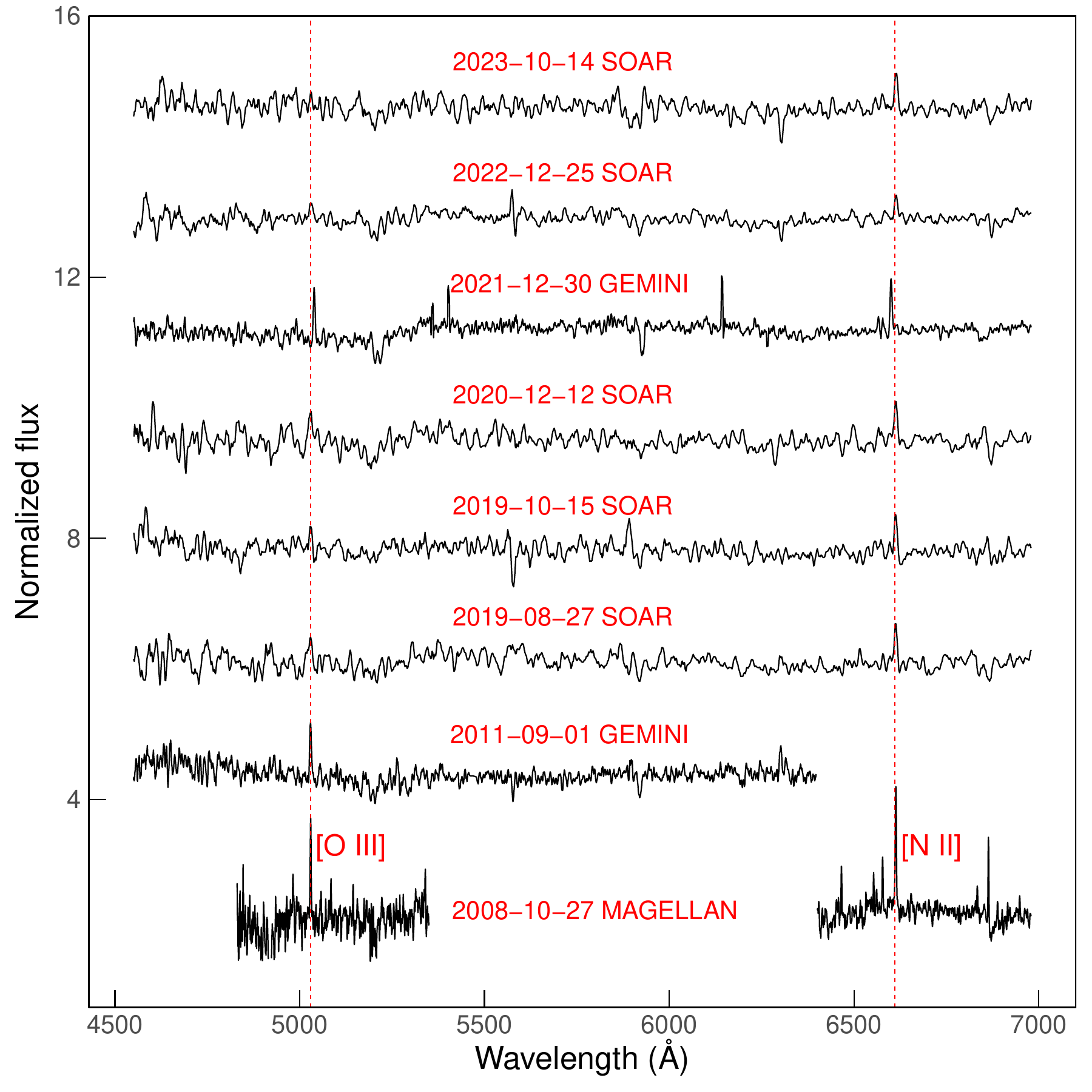}
}
\caption{(Top) Radial velocity profiles of the [N\textsc{ii}] and [O\textsc{iii}] emission lines from GCU7 from 2008 to 2023. The profiles are normalized and presented for each observation date, showing the consistency in the radial velocities of these lines. (Bottom) Evolution of the spectral emission lines of GCU7.  The spectra consistently show the presence of strong [N\textsc{ii}] and [O\textsc{iii}] emission lines, demonstrating their stability over the 15-year period.}
\label{fig:spectral_elias}
\end{figure}

GCU7 was first observed in 2006 on the 6.5-meter Magellan II Clay Telescope. In these subsequent observations, both [O\textsc{iii}] $\lambda$ 5007\AA \hspace{0.01cm} and [N\textsc{ii}] $\lambda$ 6583 \AA \hspace{0.01cm} emission lines were observed beyond the GC continuum. These lines were resolved in data taken on October 27th, 2008, with the Magellan Echellette Spectrograph (MagE), and we use this observation from \cite{Irwin} as a useful comparison point to the observations described below. We also present one Magellan Inamori-Magellan Areal Camera and Spectrograph (IMACS) spectrum of nearby GCU8. This spectrum was reduced with the same technique as described in \cite{Irwin}. We used \textsc{fxcor} \citep{tonry1979,alpaslan2009} to cross-correlate with a template spectrum of a single star, and found the radial velocity of the host cluster is 1433 $\pm$9 km/s, which is consistent with other clusters in Fornax \citep[e.g.,][]{2004AJ....127.2114D}, thus confirming that the X-ray source is indeed hosted by a GC. In \autoref{fig:spectral_elias}, we present the radial profiles (top) and spectral evolution (bottom) of the [N\textsc{ii}] and [O\textsc{iii}] emission lines from GCU7, observed from 2008 to 2023. Over the 15 years of observations, the emission lines remain consistently detectable, indicating overall stability in their presence. While some minor variations can be noted in the profiles, these are not significant enough to suggest any substantial changes in the conditions of the emitting gas. The overall persistence of the [N\textsc{ii}] and [O\textsc{iii}] lines suggests that the physical environment surrounding GCU7 has remained relatively stable over the observation period.

\subsubsection{SOAR Observations}
We observed GCU7 over 5 nights (2019-08-27, 2020-10-15, 2020-12-12, 2022-12-25, and 2023-10-14) using the Goodman High Throughput Spectrograph \citep{2004SPIE.5492..331C}, mounted on the 4.1-m Southern Astrophysical Research (SOAR) telescope in Chile. The observations were obtained using the 400\,l\,mm$^{-1}$ grating and a 0.95\, arcsec slit, providing a resolving power $R \approx 1000$ over the range. The spectra were reduced and optimally extracted using the \textsc{apall} package in the Image Reduction and Analysis Facility (IRAF; \citealt{1986SPIE..627..733T}). The wavelength calibration was performed using an iron lamp and
a flux calibration was applied.

\subsubsection{Gemini Observations} 

GCU7 was subsequently observed on 2011-09-01 with the Gemini Multi-Object Spectrograph (GMOS) on the 8.1-m Gemini South Telescope under program GS-2011B-Q-17-14. The GMOS observations were taken using the B600 grating and a 1\, arcsec long-slit, covering the range 4000\,--\,6428\,\AA \hspace{0.01cm} at a resolving power $R \approx 1700$. At this range, the [N\textsc{ii}] emission line at 6583\,\AA \hspace{0.01cm} was not covered. The same GMOS setup was used to observe GCU7 on 2021-12-30 under the program GS-2021B-FT-214, but this time at a central wavelength of 5750\,\AA \hspace{0.01cm} covering both [O\textsc{iii}] 5007\,\AA\hspace{0.01cm} and [N\textsc{ii}] 6583\,\AA \hspace{0.01cm} emission lines. We used the \textit{gemtools} package \citep{2016ascl.soft08006G} to reduce the dimensionality of the MEF files and then followed the same methodology applied to the SOAR data to extract and calibrate the spectra. 

\subsubsection{GCU7 Equivalent Width Measurement}
In all our analyses, the equivalent width of the spectra was computed following the process outlined by \cite{Dage19b}. This involved selecting and averaging regions 20 \AA \hspace{0.01cm} wide, both redward and blueward of the emission features. These regions were chosen randomly and the process was iterated 100 times to ensure robustness. A straight line was then fitted to these averages to extrapolate the cluster continuum in the absence of emission. This approach allowed us to accurately compute the equivalent width. The average equivalent width from these trials, along with the standard deviation, was reported as the measure of the equivalent width and its uncertainty, respectively. This methodology was uniformly applied to all observations and was specifically used to analyze the [O\textsc{iii}] and [N\textsc{ii}] emission features.

We also note that the equivalent widths were analyzed using a Bayesian hierarchical normal model (e.g., \cite{Willman_strader_2012}), which revealed no evidence of intrinsic variability. This suggests that the observed variations in equivalent widths are solely due to measurement uncertainties. Additionally, the continuum flux in the spectrum represents the integrated light of all stars in the GC and is therefore not expected to vary.

\begin{figure}
\centering
\subfigure{
\includegraphics[width=0.49\columnwidth]{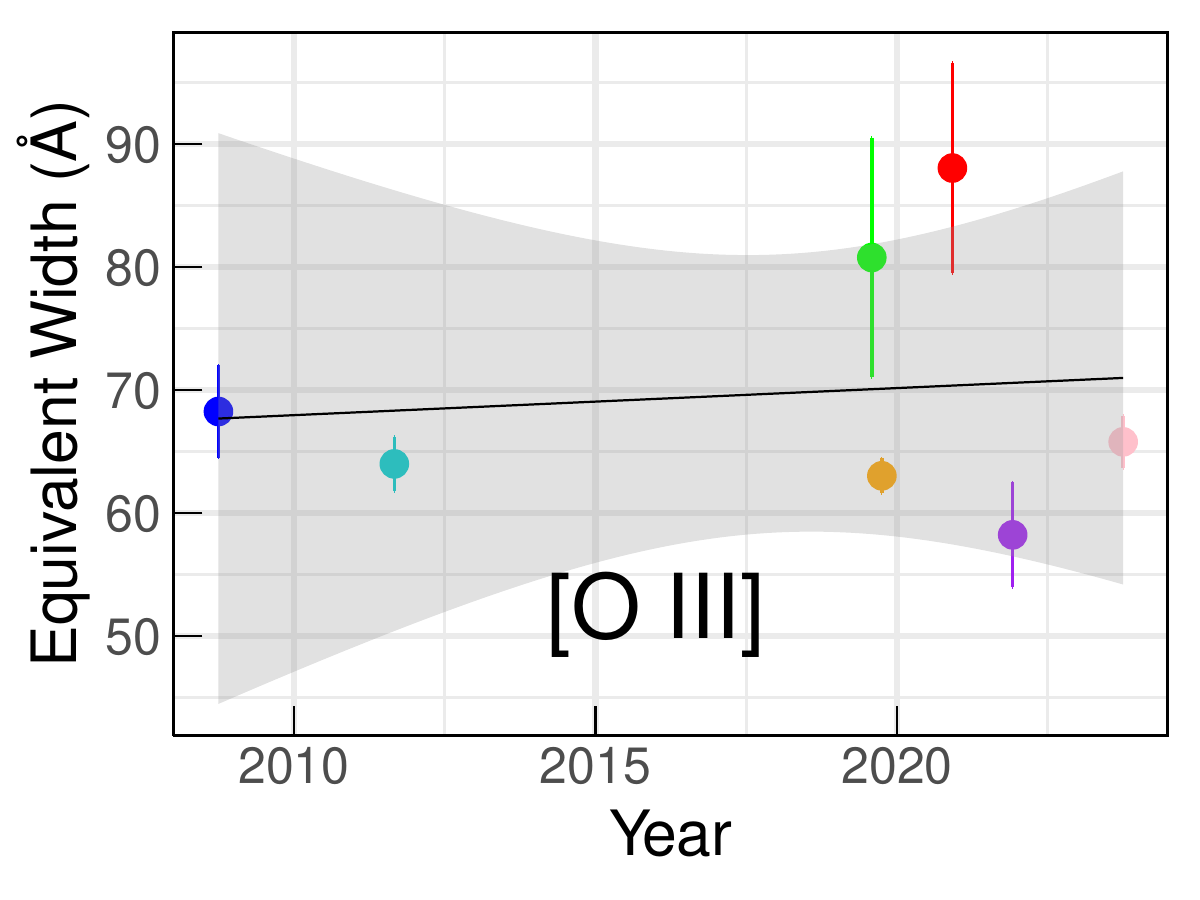}
\includegraphics[width=0.49\columnwidth]{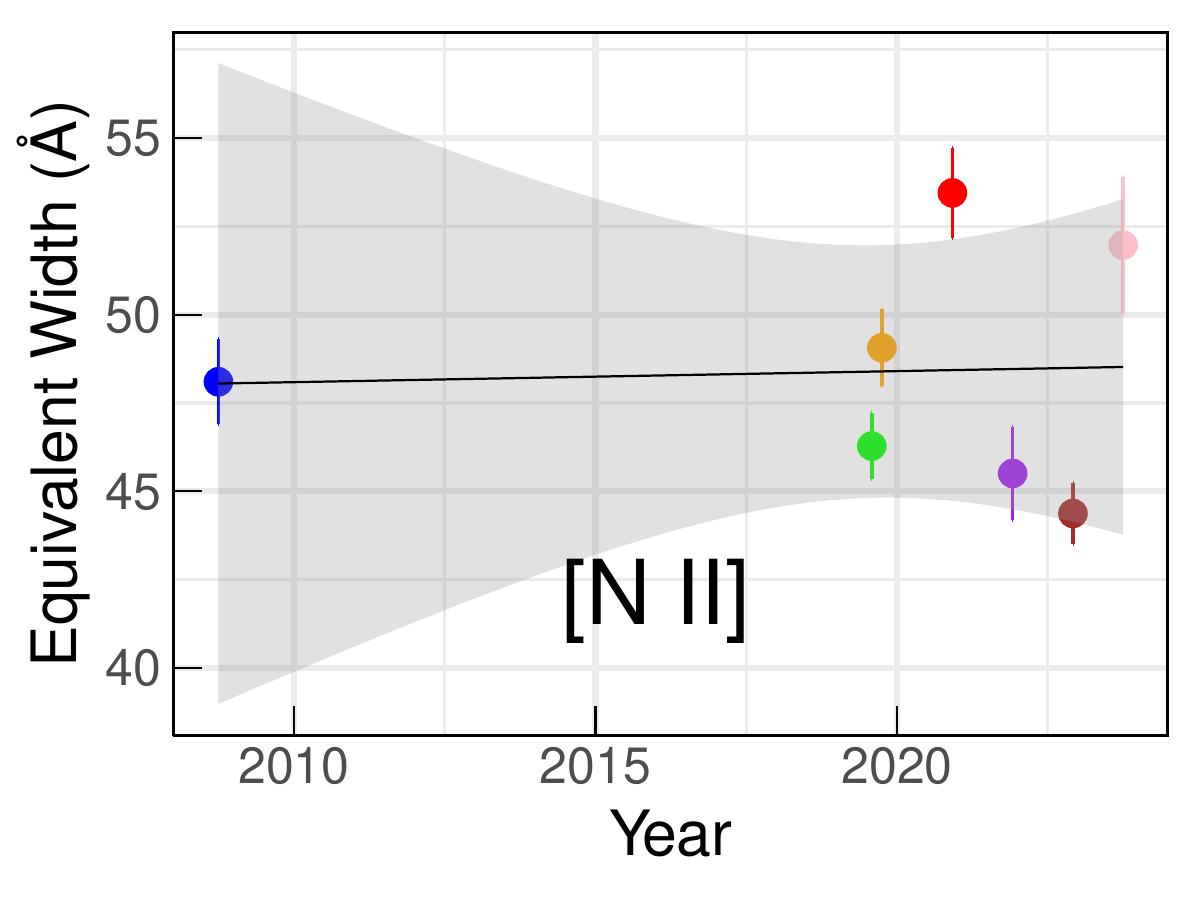}
}
\subfigure{
\includegraphics[width=\columnwidth]{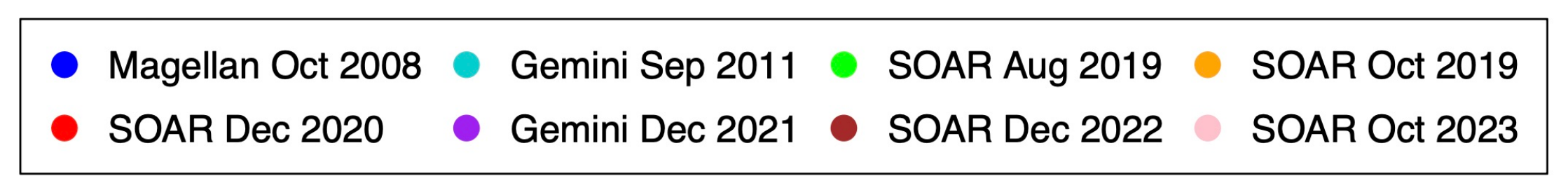}
}
\caption{An evolutional equivalent widths trend between different observations over time. The left panel shows the [O\textsc{iii}] emission line and the right panel shows the [N\textsc{ii}] emission line. The error bars represent the uncertainty in the EW measurements, and the shaded region indicates the 95\% confidence interval. The detailed measurements are presented in \autoref{table:eqwid}.}
\label{fig:EW}
\end{figure}

\autoref{fig:EW} presents the optical spectra of GCU7, illustrating the process used to calculate the equivalent widths of the emission features. The detailed equivalent width measurements are provided in \autoref{table:eqwid}. This figure highlights the evolutionary trend in equivalent widths over different observational periods, with the left panel displaying the [O\textsc{iii}] emission line and the right panel displaying the [N\textsc{ii}] emission line. The error bars represent the uncertainties in the equivalent width measurements, and the shaded regions indicate the 95\% confidence intervals.

\section{Results and Discussion}
\label{sec:discussion}
We present new and archival X-ray and optical spectroscopic observations of two GC ULXs in NGC 1399. We briefly speculate on the nature of GCU8, the brightest GC ULX, best fit by a hard power-law emission model, with no evidence for X-ray variability, and hosted by a metal-rich GC with no evidence for optical emission beyond the cluster continuum. We then go on to discuss new observations of GCU7 in light of our current understanding of population synthesis, whether it can still be described as an IMBH TDE, and whether it can be suitably explained as an UCXB. 

\subsection{The Nature of GCU8} 
\begin{figure}
    \centering
    \includegraphics[width=0.9\columnwidth]{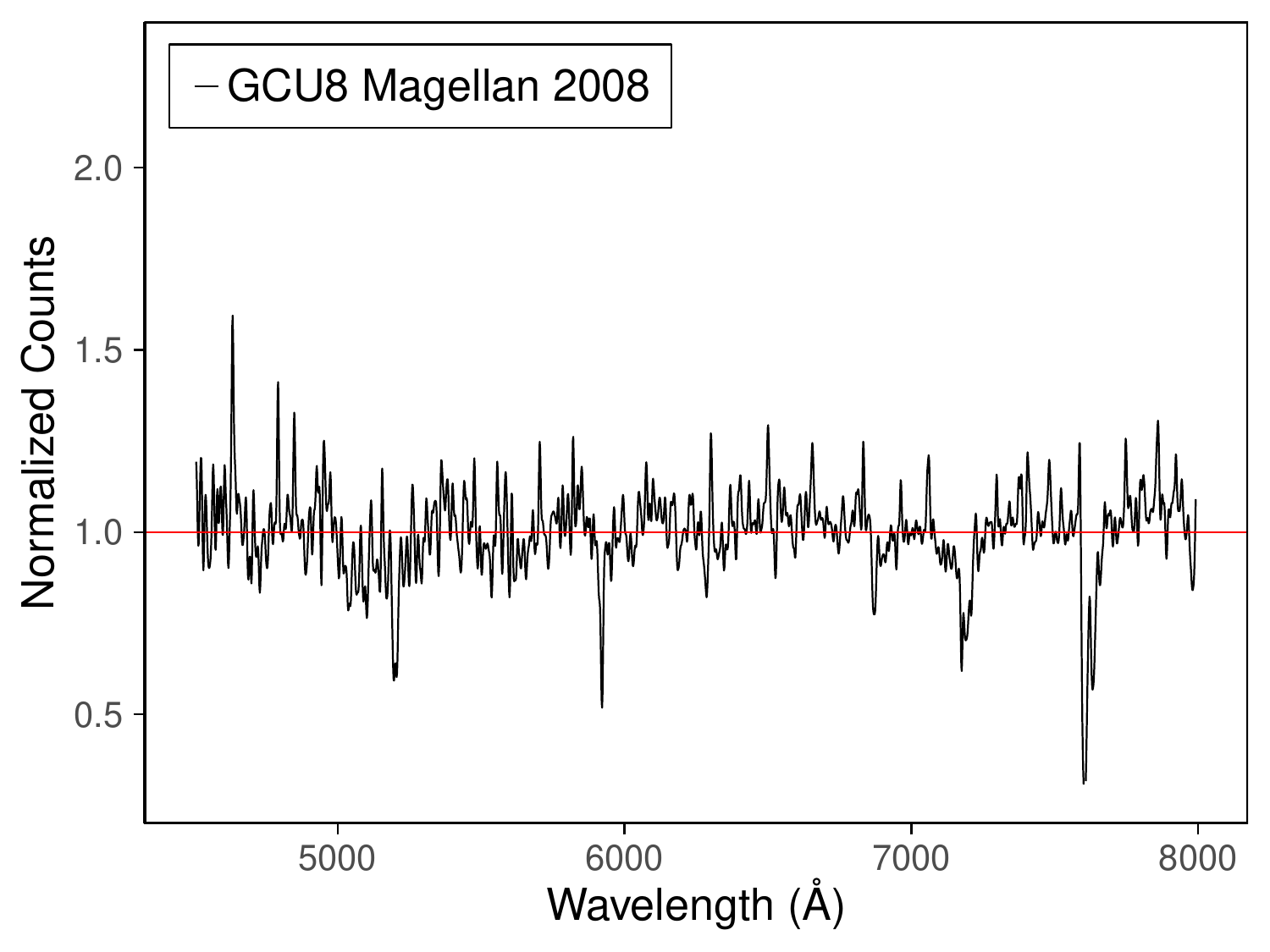}
    \caption{Magellan IMACS spectrum of GCU8 from Nov 26th 2006. No optical emission is detected above the cluster continuum.}
    \label{fig:GCU8}
\end{figure}

The source GCU8 is the brightest of all the observed GC ULXs, reaching a peak X-ray luminosity over 4.8 $\times 10^{39}$ erg/s. The host cluster is metal-rich with $g-z$=1.6 \citep{jordan_2015}, and, as described in Section 2.4, the cluster radial velocity is 1433 $\pm$9 km/s, confirming the cluster nature of the source. There is no sign of any optical emission beyond the cluster continuum (\autoref{fig:GCU8}, and the source does not vary significantly in X-ray \citep{Dage19a}. The X-ray spectrum of the source is rather hard, cannot be fit with a thermal component, and the average power-law index from \cite{Dage19a} is $\Gamma$= 1.3. Based on \cite{Sutton_2013}, GCU8 would fit within the high-ultraluminous regime. Studies such as \cite{Gladstone_2009} and \cite{Walton_2018} further support this, showing that ULXs typically exhibit a thermal continuum from a broadened accretion disk, with a hard excess. As shown in \autoref{tab:myfits}, GCU8 is still extremely bright, and the best-fit power-law slope is consistent with uncertainties to previous measurements from \cite{Dage19a}. 

\cite{2007ApJ...662..525K}'s study of low mass X-ray binaries (LMXBs) in GCs hosted by NGC 1399 calculates a 24\% chance that a GC can host 3 LMXBs, and that the most metal-rich clusters in NGC 1399 should show a high, but not variable, X-ray luminosity due to LMXB superposition. While it is thus possible that GCU8 may be due to the superposition of many LMXBs, our understanding of metallicity and the role it plays in GC environments is not sufficient yet to draw a satisfactory conclusion about the nature of GCU8.

\subsection{GCU7 and Tidal Disruption Events}

The model of a tidal disruption by an IMBH from \cite{2011ApJ...726...34C} can be ruled out based on the longevity of the [O\textsc{iii}] emission; The [O\textsc{iii}] emission has been observed for 17 years (or 6200 days), it exceeds the prediction from Figure 5 of \cite{2011ApJ...726...34C} which shows that the [O\textsc{iii}] emission should show a significant decay beginning after 1000 days, dropping by a factor of a few after 3000 days. Instead, we are seeing remarkably steady behaviour in the emission lines, particularly from 2008--2022, with any marked differences being minimal enough that they are more likely due to instrumental effects/instrumental differences. 

Furthermore, the longevity of the X-ray observations (over 24 years including ROSAT observations) makes it difficult to understand how to apply leading models for disruption by an IMBH more generally, as the \textit{Chandra} observations of GCU7 remain steadily bright over a long enough time scale to rule out any models that require short-term variability, such as partial TDEs \citep[e.g., those identified by][among others]{2023arXiv231003782S}.

While stable spectral lines of GCU7 strongly argue against a TDE scenario, we note that not all TDEs show canonical decay patterns, and some cases show irregular behaviour in X-rays \citep{Mummery_2024}. In rare cases, repeating partial TDEs, such as the X-ray source J0456-20, have been observed to cycle through distinct phases of high and low luminosity \citep{Liu_2023}. However, we do not see those events in GCU7.

\subsection{Modeling the Emission lines of GCU7}
\label{sec:modeling}
BH + CO WD binaries can form UCXBs \citep{2017ApJ...851L...4C, 2012A&A...537A.104V}, although such UCXBs likely resemble RZ 2109, as discussed previously. BH + He WD binaries can also form UCXBs, although such binaries are much less common than neutron star (NS) + He WD UCXBs at Solar-like metallicities, based on Galactic observations. Additionally, in evolved BH + He WD systems, when the mass ratio becomes sufficiently extreme, the WD donor can be overwhelmed by the tidal forces exerted by the disc around the much more massive BH accretor, potentially leading to unstable mass transfer or even complete disruption of the donor \citep{lasoata_2008, Yungelson_2008}. 

Given the observed stability of GCU7's X-ray luminosity over two decades and the persistent optical emission lines, it is more plausible that GCU7 is a UCXB consisting of a NS primary accreting from a He WD donor. NS + He WD systems can sustain stable mass transfer over long timescales \citep{nelemans_2010}. The relatively small mass ratio in NS + He WD systems is conducive to stable mass transfer \citep{2017MNRAS.467.3556B}.

The formation of NS + He WD UCXBs in a dense environment like GC largely has two different scenarios. One plausible channel of the formation would be through direct collision between a NS and an evolved star, such as a red giant or an evolved subgiant. 
In this scenario, when the NS collides with an evolved star, a close encounter can strip the envelope of the evolved star, leaving behind its helium core \citep{Verbunt1987}. This tight orbit of the system allows the He WD to fill its Roche lobe, initiating stable mass transfer onto the NS to form UCXB.

On the other hand, exchange interactions can also result in the formation of NS + He WD binaries. During such dynamical interactions, the NS captures the He WD and exchanges its original companion star, forming a new binary system. However, this scenario typically involves longer timescales, as the newly formed binary must undergo orbital decay through gravitational radiation before the mass transfer can occur \citep{Ivanova2008}.

\begin{figure*}
\centering
\subfigure{
\includegraphics[width=0.90\textwidth]{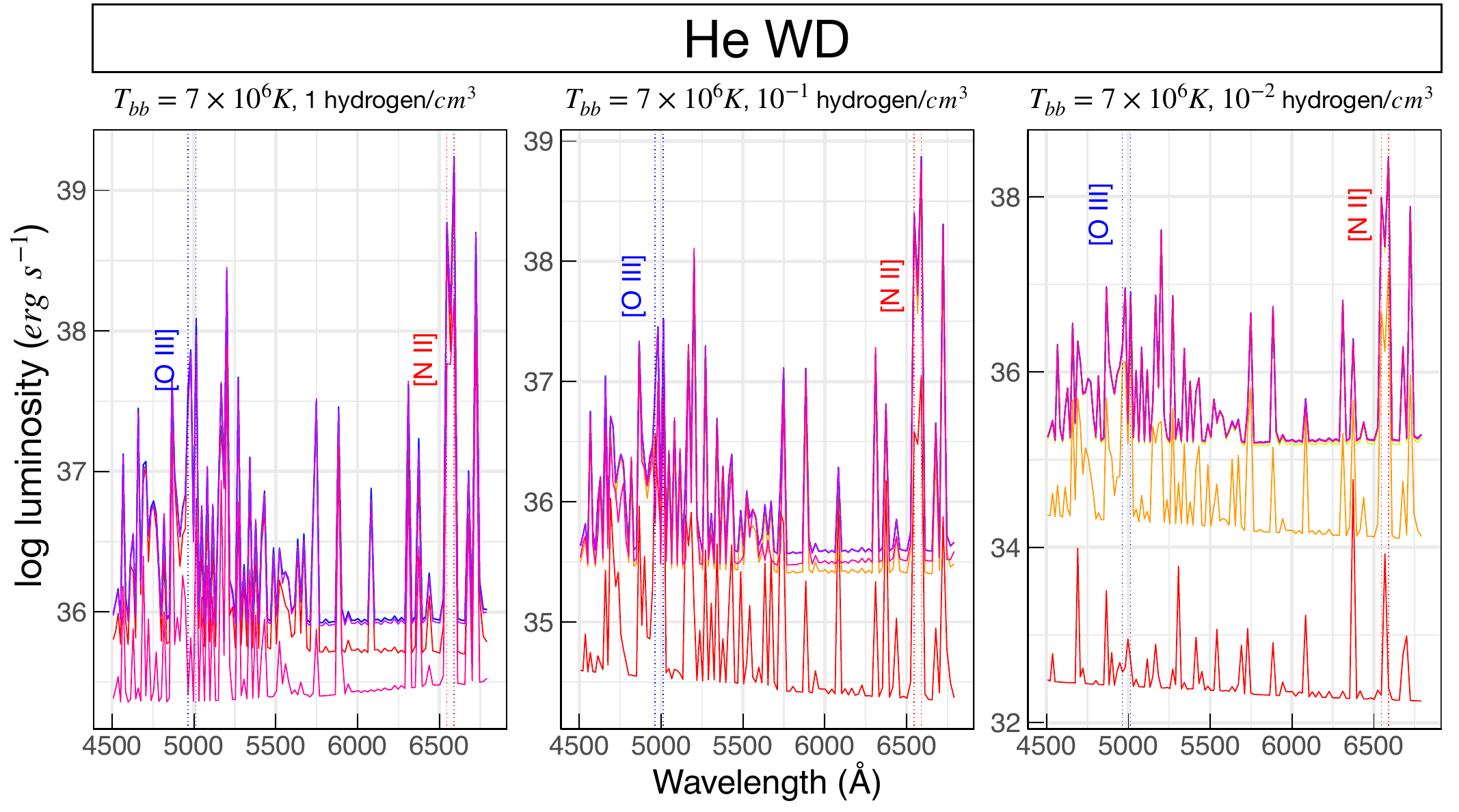}
}
\subfigure{
\includegraphics[width=0.89\textwidth]{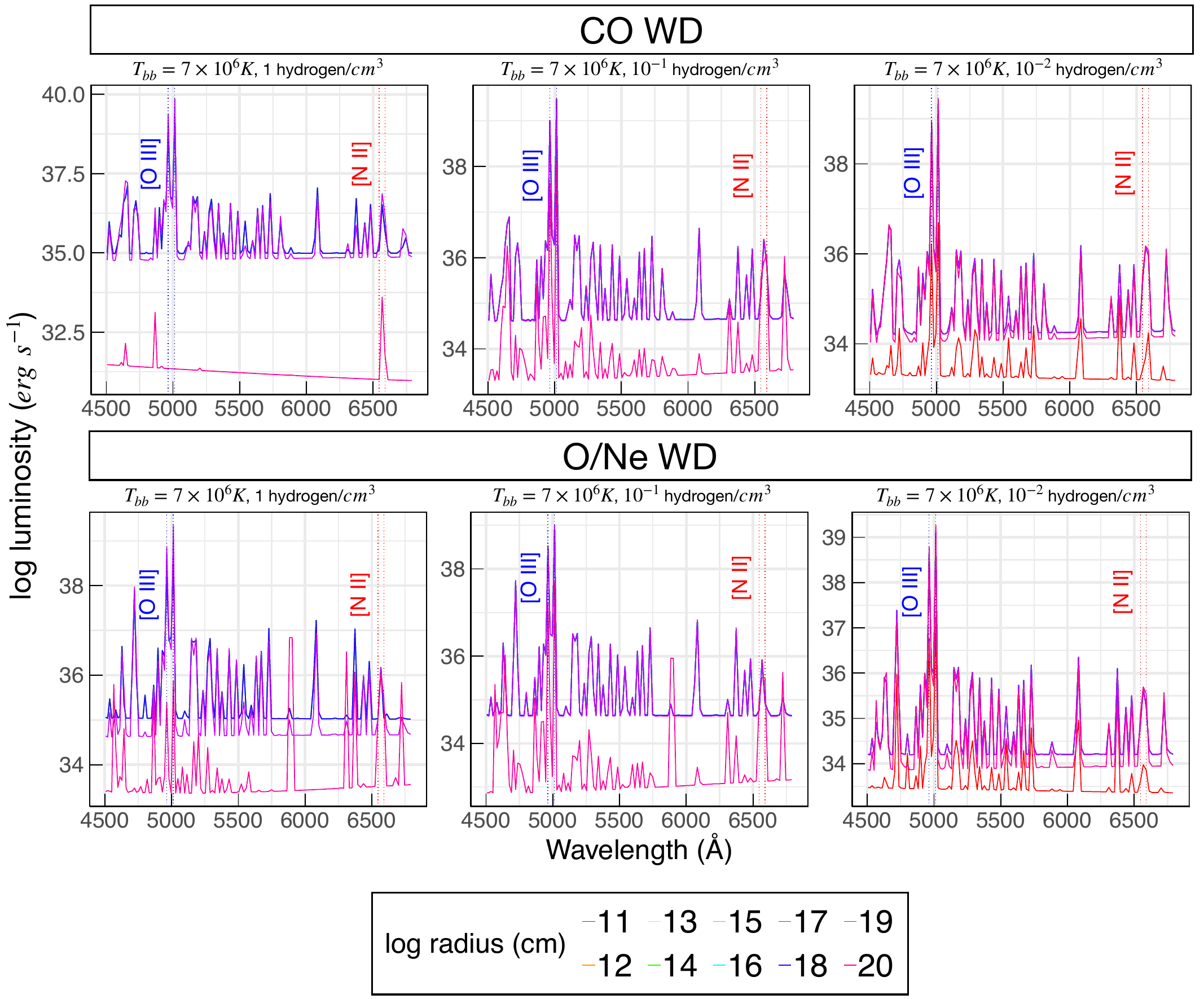}
}
\caption{Simulated spectra for three different WD composition models: He WD (top), CO WD (middle), and O/Ne WD (bottom). Each model explores varying hydrogen densities ($n_{\text{H}} = 0.1, 10^{-1}, 10^{-2} \, \text{cm}^{-3}$) with a constant blackbody temperature of $T_{\text{bb}} = 7\times10^6 \, \text{K}$. The x-axis represents the wavelength range in Angstroms (Å). The y-axis indicates the logarithmic luminosity of the system, measured in erg/s. The colour of each line represents the inner radius of the gas from the source, on a logarithmic scale.}
\label{fig:CONeWD}
\end{figure*}

In a NS + He WD UCXB, mass transfer is driven by gravitational wave radiation, which shrinks the orbit and brings the components closer together \citep{2004ApJ...607L.119B}. As the He WD fills its Roche lobe, mass transfer commences and can proceed stably due to the favorable mass ratio \citep{Tauris_2012,2017MNRAS.467.3556B}. Early in their evolution, these systems can exhibit high mass-transfer rates that approach or exceed the Eddington limit, leading to high X-ray luminosities. Additionally, recent studies of field ULXs, such as NGC 247 ULX-1, have identified helium donors in systems powered by super-Eddington accretion (e.g., \cite{Dai_2021,Zhou_2023}). Such mass transfer is driven by a combination of gravitational wave emission and mass loss, as well as stability of the accretion disk \citep{Heinke_2013}. 

To further investigate the nature of GCU7 and its emission properties, we performed spectral simulations using the large-scale spectral synthesis code \textsc{CLOUDY} \citep{cloudy_1998}, modeling a photoionized plasma. This code predicts the emitted spectrum by simulating the physical conditions inside such an astronomical medium.

In our simulation, we adopted a blackbody temperature of $7\times10^6$K, which corresponds to 0.57$\pm{0.09}$ keV from the X-ray spectral fitting of the disk blackbody model. We used a source distance of 21.1 Mpc and assumed an inner accretion disk radius of approximately $10^6$ cm. Although we varied the temperature according to the blackbody fitting results, which provided a range of approximately $5.5 \times 10^6$K to $7.6 \times 10^6$ K, we chose $7\times10^6$K for this study.

In terms of abundances, we assume that nitrogen is enhanced relative to oxygen by at least an order and up to two orders of magnitude above solar, corresponding to a N/O ratio of roughly 20 times solar abundances. We emphasize that this elevated ratio is not derived from a formal fit; rather, it is a plausible approximation consistent with the observed strengths of [N\textsc{ii}] and [O\textsc{iii}] and the lack of hydrogen lines.

We explored which WD abundances could produce the observed strong emission lines of [N\textsc{ii}] and [O\textsc{iii}] by utilizing the abundances for three different types of WDs (He WD, CO WD, O/Ne WD) from Table 1 in \citet{koliopanos_2013}. We assumed a hydrogen density of $n_{\mathrm{H}} = 0.1$ hydrogen $\mathrm{cm^{-3}}$, reflecting typical conditions in GC environments, but also varied the hydrogen density from 1 hydrogen $\mathrm{cm^{-3}}$ to $10^{-2}$ hydrogen $\mathrm{cm^{-3}}$ to account for different cluster conditions. 
\citet{Freire_2001} estimated an electron density of $n_e= 0.67 \pm 0.15 \, \mathrm{cm}^{-3}$ in the core region of 47 Tucanae based on pulsar dispersion measures, providing a robust reference for the ionized gas density in GCs. Furthermore, \citet{Naiman_2019} estimated that under specific conditions, the gas density in GC environments can reach $\sim 0.1 \mathrm{cm^{-3}}$, consistent with our adopted density.
Since we modeled the emission as dominated by an accretion disk exceeding the Eddington limit, we set a column density of $N_{\mathrm{H}} = 10^{22}\mathrm{cm^{-2}}$. This value is consistent with the column density found in the X-ray emitting and absorbing plasma of NGC 1313 ULX-1 \citep{Pinto_2020}. Additionally, we assumed a spherical geometry for the emitting region, which simplifies the modeling process and is reasonable for modeling extended outflows from super-Eddington accretion disks.

\autoref{fig:CONeWD} shows the simulated spectra for the different WD compositions: He WD, CO WD, and O/Ne WD. We varied the hydrogen density while keeping the temperature constant to examine whether the system would still produce strong [N\textsc{ii}] and [O\textsc{iii}] emission lines at different inner radii. The x-axis represents the wavelength range in Angstroms (Å), and the y-axis indicates the logarithmic luminosity in erg/s. The inner radius of the gas is represented by the colour of the lines, shown in a logarithmic scale (cm). In the top panel, the He WD abundance model shows both [N\textsc{ii}] and [O\textsc{iii}] emission lines, whereas the CO WD and O/Ne WD models primarily showed strong [O\textsc{iii}] emission with weaker [N\textsc{ii}] lines. 

\begin{figure}
\centering
\subfigure{
\includegraphics[width=0.98\columnwidth]{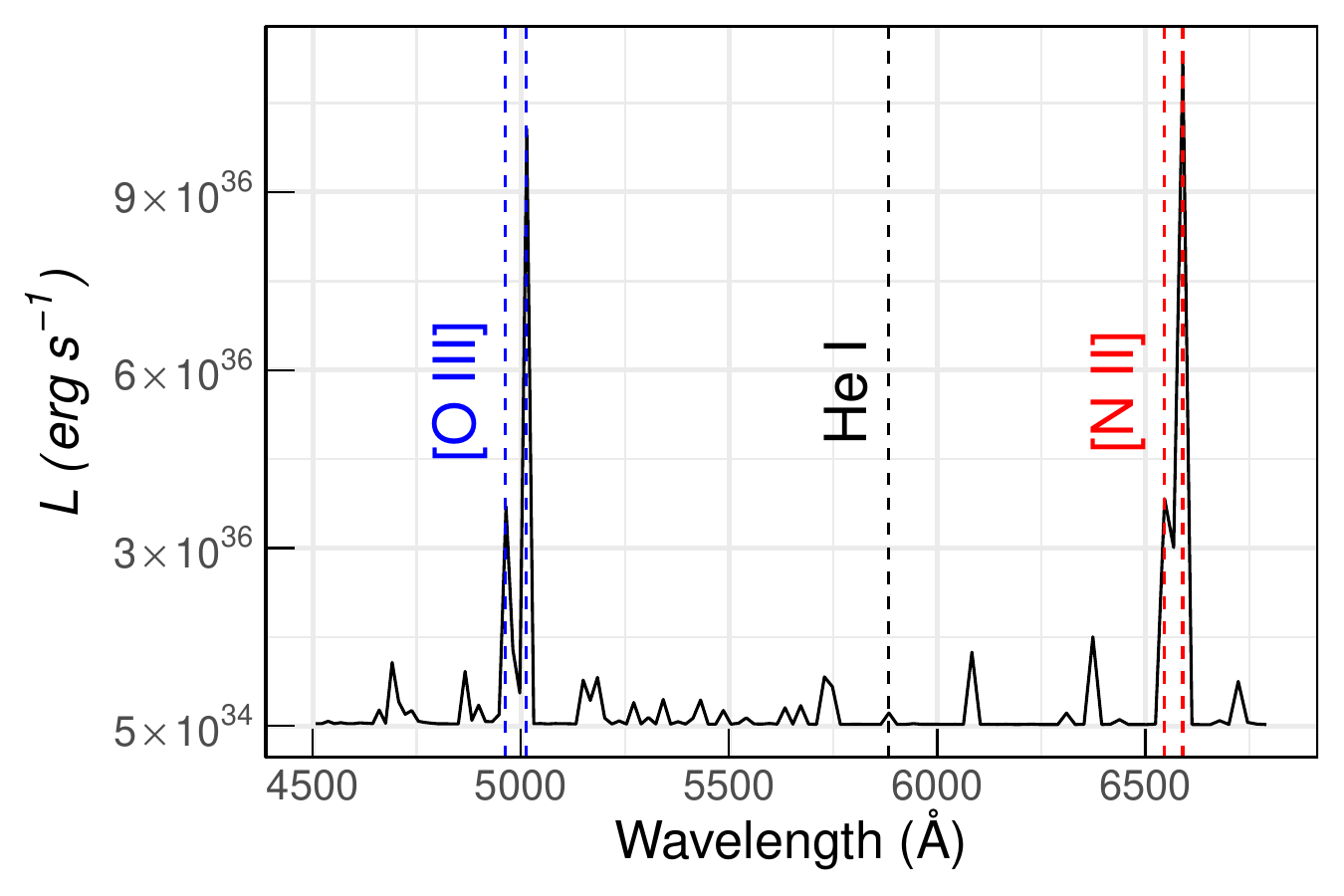}
}
\caption{The \textit{Cloudy} spectral emission lines from the He WD show strong [O\textsc{iii}] and [N\textsc{ii}] lines. The model uses a blackbody temperature of $7\times10^6$ K, hydrogen density of $0.1$ hydrogen cm$^{-3}$, and an inner gas radius of $10^{11}$ cm.}
\label{fig:cloudy}
\end{figure}

Based on the \textit{CLOUDY} result, we investigate to find the best model in He WD abundances which corresponds to the observed emission lines. The He WD abundances showed the best results when the inner radius of the gas was set at $10^{11} \mathrm{cm}$. \autoref{fig:cloudy} shows the spectral emission lines and the luminosities of the emission line of each [N\textsc{ii}] and [O\textsc{iii}] well correspond with the observed luminosity while showing a lack of helium lines. 

In the ULX Holmberg II ULX-1, \cite{Barra_2024} also found evidence for enhanced nitrogen with respect to solar values. However, in that case, the source is from a young stellar population with a massive star donor. Since this GC ULX is from an old stellar population, the origin of the enhanced N may be different.

The narrowness of the optical emission lines (FWHM $\sim 140 km/s$) suggests they are unlikely to have formed in a disk wind or outflow very close to the compact object. However, the presence of [N\textsc{ii}] and [O\textsc{iii}] already indicates that the emission primarily originates from a lower-density region, perhaps a spatially extended nebula. Photoionized nebulae are observed around some other ULXs and have low velocities \citep{Pakull2002, Kaaret2009}. Most ULXs are in star-forming galaxies with significant ISM surrounding the source and, hence, are not exact analogues to GCU7, but the general idea of the optical emission primarily coming from photoionized nebulae seems plausible given the existing data. Deeper future optical spectroscopy and high-resolution narrow-band imaging could better constrain the presence of a fainter, broader component that might be formed closer to the binary and the existence of an extended nebula.

The production of strong [N\textsc{ii}] and [O\textsc{iii}] lines can be attributed to the following physical processes:
\begin{itemize}
    \item Super-Eddington Accretion and Outflows: In the early stages of a NS + He WD UCXB, the mass transfer rate can exceed the Eddington limit for a NS ($\dot{M}_{\mathrm{Edd}} \approx 4 \times 10^{-8} M_{\odot}~\mathrm{yr}^{-1}$). This leads to radiation-driven outflows that carry processed material from the donor into the surrounding environment.
    
    \item Photoionization of Outflowing Material: The intense X-ray emission from the accretion disk photoionizes the outflowing material, resulting in emission lines from ions such as N$^{+}$ and O$^{++}$ 

    \item Chemical Composition of the Donor: The donor's nitrogen- and oxygen-enriched, carbon-depleted outer layers—resulting from its evolutionary history—amplify the corresponding emission lines when this material is transferred to the NS.
\end{itemize}

The lack of H points most directly to a WD or helium star as the donor, and the old age of the cluster strongly suggests it is the former.
The apparent enhancement of N in the spectrum suggests CNO processed material. It is relevant to note here that despite the domination of energy generation in stars $\sim1M_{\odot}$  and lower by the pp process, the CNO process also operates in the core of a solar-mass main sequence star, especially as the central H is depleted \citep{Iben1967}. This means that solar-type stars turn nearly all of their initial C and some of their initial O into N, such that the He core of such a star early on the giant branch is about 1\% N by mass, with less O and almost no C. While additional CNO-cycle shell burning would also occur on the giant branch, this would not meaningfully change the relative abundances. If the envelope of such a star were stripped, it would leave behind a He WD with enhanced N, some O, and strongly depleted C.

While this scenario assumes solar composition, the location of this star in a GC opens up a plausible or even likely additional possibility. The majority of the stars in Galactic GCs show unusual abundances that indicate light-element self-enrichment, with a larger fraction of self-enriched stars for higher cluster masses \citep{Carretta2010}. While there is a spread among the stars, typical stars would show strongly enhanced N ([N/Fe] = +1 or higher), mildly enhanced O ([O/Fe] = +0.2), and depleted C ([C/Fe] = --0.2; see the compilation of \citealt{Roediger2014}). The origin of this material is not yet known but involves H-burning at high temperatures in some manner \citep{Bastian2018}. In any case, there is evidence that extragalactic GCs show similar self-enrichment (e.g., \citealt{Schiavon2013,Peacock2017}), implying that a randomly chosen star in a massive cluster such as GCU7 would be likely to show similar anomalies. In this case, the He WD resulting from the same scenario would have even higher N (potentially up to 2\% by mass), about 1\% O, and still trace C. The implication is that strongly N-enhanced material, with some O and highly-depleted C, is the expected composition of an outflow from a high $\dot{M}$ GC UCXB with a He WD donor, and no unusual assumptions are necessary to reach this prediction. Future UV observations of this source could constrain the abundance of C and test this scenario.

The stability of the mass transfer in the NS + He WD system ensures that these processes can persist over long timescales, consistent with the two decades of observations for GCU7. Additionally, the absence of significant variability in the X-ray luminosity aligns with expectations for stable mass transfer in such systems.

\subsection{Evolutionary Modeling of GCU7 as a Young UCXB}
\begin{figure}
    \centering
    \includegraphics[width=0.9\columnwidth]{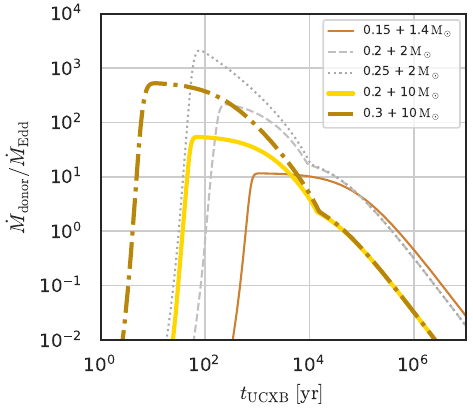}
    \caption{Evolutionary tracks for UCXB systems consisting of a He WD donor and NS/BH accretors. All systems reach highly super-Eddington rates early in their evolution, subsequently reaching rates similar to the ones observed in GCU7 and later approaching the low rates seen in older Galactic UCXBs.}
    \label{fig:UCXBTracks}
\end{figure}

Here, we consider the likely evolutionary history of GCU7 assuming it is a young UCXB. In Figure~\ref{fig:UCXBTracks}, we show long-term stellar evolution tracks for several plausible progenitors undergoing stable mass transfer: a $0.15\,\text{M}_\odot$ He WD donor with a regular $1.4\,\text{M}_\odot$ NS, more massive $0.2\,\text{M}_\odot$ and $0.25\,\text{M}_\odot$ He WDs donors with a more massive $2\,\text{M}_\odot$ NS, and two higher-mass, $0.25\,\text{M}_\odot$ and $0.3\,\text{M}_\odot$ He WDs with a $10\,\text{M}_\odot$ BH companion. We construct the figure based on the secular evolution code from \cite{2017MNRAS.467.3556B}, \cite{2017ApJ...851L...4C}, assuming mass transfer to be conservative for $\dot{M}\lesssim 10^{-6}\,\text{M}_\odot$ and proceed through disk winds at rates $\dot{M}> 10^{-6}\,\text{M}_\odot$, following equation 10 in \citet{2017MNRAS.467.3556B}. 

We can see that all UCXB systems experience a phase of super-Eddington mass transfer and reach the rates of about $10\dot{M}_{\text{Edd}}$ at ages between $10^3\,\text{yr}$ and few $10^4\,\text{yr}$. We can also observe that UCXB tracks typically firstly pass highly super-Eddington phase resembling RZ2109 \citep{2024MNRAS.529.1347D}, then later reach states similar to those in GCU7 and then, much later, reach the rates of Galactic sources such as 4U 1820-30. We also note that UCXB models predict no measurable evolution in the observed mass transfer rates over a $20$ year span. Finally, systems like RZ 2019 and GCU7, given their young age, can uniquely constrain mass transfer models for UCXB evolution.

Distinct evolutionary stages in the UCXB scenario are likely to produce different outflow geometries \citep{2017MNRAS.467.3556B}. The youngest UCXBs, potentially such as GCU7, with the highest $\dot{M}/\dot{M}_{\text{Edd}}$ are likely to produce mechanically pushed slow, dense wind, explaining the narrowness of the lines. Super-Eddington systems with lower $\dot{M}/\dot{M}_{\text{Edd}}$, such as RZ 2109, have less dense mechanical wind, such that the fast bipolar jetted outflow can penetrate and escape through it, thus producing a combination of broad lines from the jet and narrow lines from the slow wind  \citep{2024MNRAS.529.1347D}. Near-Eddington and sub-Eddington UCXBs, such as the Galactic UCXBs, may launch outflows during outbursts.

\subsection{UCXB Population via Monte-Carlo simulation}
To gain insights into the formation and evolution of UCXBs, we utilized simulations from the Monte Carlo cluster simulator \citep[MOCCA;][]{1998MNRAS.298.1239G} and integrated the classifications and data from \cite{Oh_2024}. They divided the GCs into three different classes based on the core density evolution on each simulated cluster at 12 Gyrs which reflects the present-day population (PDP): Class I (expanding systems), Class II (collapsing systems), and Class III (post-core collapse or re-expansion).

\begin{figure}
\centering
\subfigure{
\includegraphics[width=0.9\columnwidth]{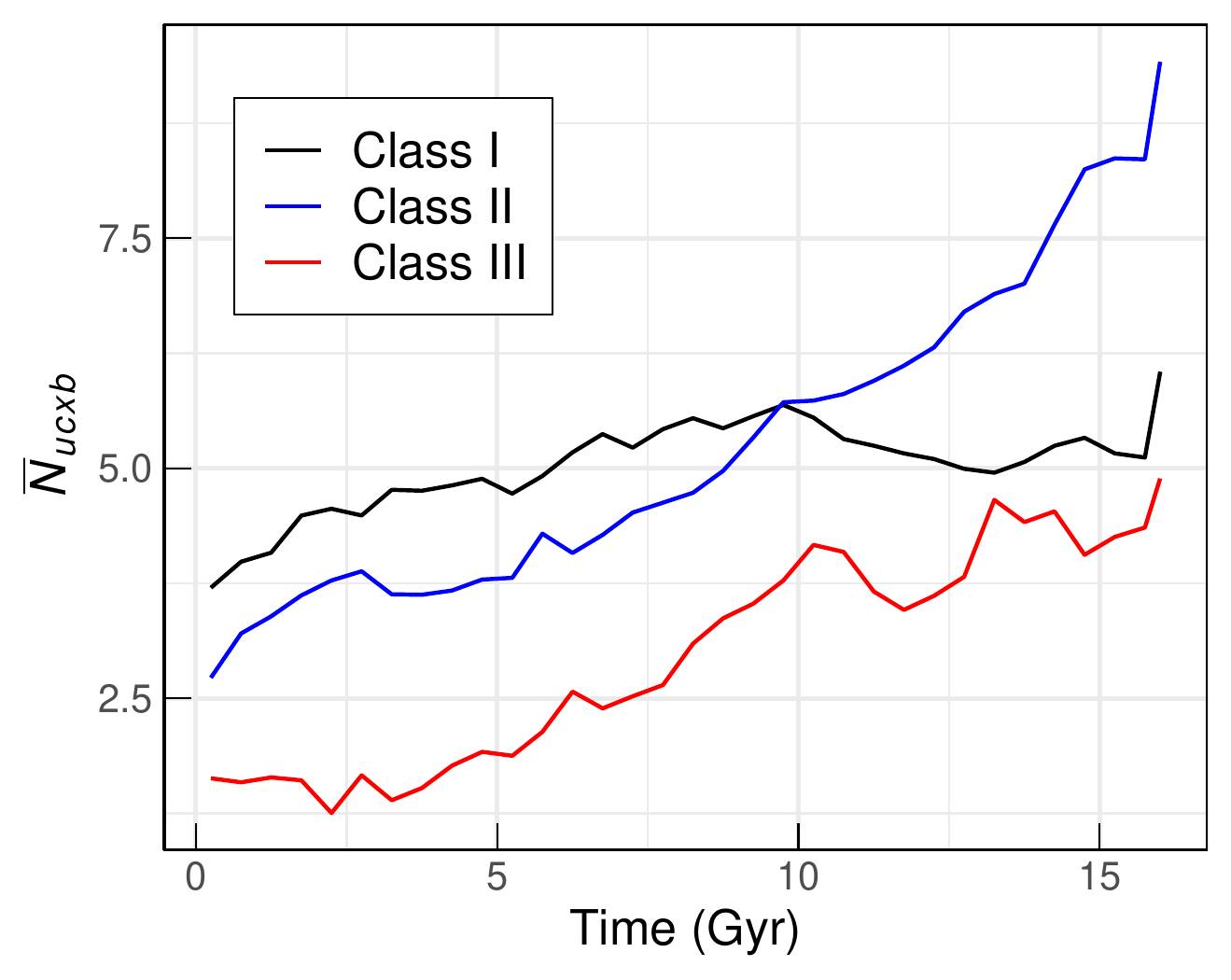}
}
\caption{The averaged number of UCXB candidates over time of the MOCCA simulation. The highest production rate of UCXBs is found in class II clusters, indicating the importance of the dynamical status of GCs for the formation of compact binaries.}
\label{fig:mocca}
\end{figure}

We identified potential UCXB candidates within these simulated clusters, focusing on WD + NS/BH binaries. \autoref{fig:mocca} shows the variations in UCXB populations across different dynamical states of GCs in MOCCA. In particular, Class II clusters, identified by their collapsing state, showed the highest production rate of UCXBs at PDP. The higher stellar densities in these clusters are responsible for the high production rate because of the frequent dynamical interactions. Additionally, \cite{Wiktorowicz_2025} investigated ULX populations using MOCCA, emphasizing the critical role of GC dynamics in their formation. Their findings underscore the significance of frequent stellar interactions and binary hardening, processes that align well with the conditions in denser clusters in our simulations. Their results indicate that NS ULXs represent a smaller fraction of in-cluster systems ($\sim$4\%).  GCU7's stable X-ray luminosity and persistent [NII] and [OIII] emission lines may be consistent with such rare NS systems, highlighting the diverse outcomes of dynamical interactions in GCs.

This section highlights that UCXB formation likelihood depends on specific core density conditions in GCs, but observed UCXBs do not necessarily reside in clusters currently exhibiting such conditions. Additionally, the term 'young UCXB' refers exclusively to systems in the early stages of mass transfer and is unrelated to the dynamical state of their host clusters. These results remain preliminary and serve as a starting point for further investigation. Our filtering criteria were based on the basic nature of UCXBs, specifically focusing on WD + NS/BH binaries as potential candidates. Despite these limitations, our approach provides a useful framework for identifying and understanding the potential formation pathways of UCXBs in GCs.

\section{Summary \& Conclusion}
\label{sec:summary}
In this study, we examined the spectral properties and evolutionary paths of ULX GCU7 in the NGC 1399. We used over 20 years of X-ray data from \textit{Chandra} and 15 years of optical spectra from Gemini, SOAR, and Magellan to analyze GCU7. Our primary observations include persistent [N\textsc{ii}] and [O\textsc{iii}] emission lines and stable X-ray luminosity over two decades. We posit that GCU7 is a young UCXB with a He WD donor, supported by theoretical insights on helium dynamics in UCXB systems. We note that the most similar GCULX source, RZ2109, with a CO WD companion \citep{zepf08, 2014ApJ...785..147S},  exhibits broad [OIII] emission lines with significant variations in equivalent width over timescales of many years \citep{Dage19b}, and shows strong X-ray variability on the scale of a few hours \citep{2007Natur.445..183M, 2024MNRAS.529.1347D}. While both sources are ULXs hosted by extragalactic GCs, with forbidden emission lines, their time-domain behaviour is quite different.

Both GCU7 and RZ2109 can be reasonably explained by common dynamical processes that occur in GCs, which likely lead to the formation of UCXBs. For GCU7, a He WD donor appears to be the most probable companion, while RZ2109 can be explained with a CO WD donor. These systems represent short-lived but potentially predictable phases in the evolution of dynamical binary interactions within dense stellar environments. Indeed, both theoretical arguments and direct observations of extended ionized bubbles around ULXs suggest that super-Eddington accretion episodes often last on the order of $10^6$ yr \citep{Pakull2002,pakull_2003}. However, in the gas-poor environments of GCs, ULX evolution may be governed by different mechanisms.

We also present archival optical spectroscopy for the nearby GCULX, GCU8, confirming it as a GC with no significant emission features and stable long-term X-ray properties. GCU8 remains the most X-ray luminous GC ULX known to date.

Our main findings are summarized as follows: 
\begin{itemize}
    \item It is unclear whether GCU8 is an individual LMXB source or a superposition of multiple LMXBs, highlighting the need to understand metallicity's role in the dynamical formation of compact binaries in GCs.
    \item GCU7's long-term multiwavelength nature is consistent with observations of other UCXBs.
    \item The \textit{CLOUDY} modeling demonstrated that helium-rich donors may not produce significant helium emission detectable at extragalactic distances, unlike oxygen and nitrogen which are abundantly produced.
    \item Our results show that if GCU7 is indeed a UCXB, it provides a promising connection to test and constrain leading theoretical simulations of GCs and the models of the early evolution of young UCXBs.
\end{itemize}

Long-term multiwavelength studies for these unique X-ray sources and their environments are essential for understanding their natures. Our findings demonstrate how state-of-the-art GC simulations can be compared to extragalactic systems, providing an exciting step for future research. By comparing production rates of UCXBs across different dynamical states of GCs, we can bridge theoretical models and observational data, despite the inherent challenges and assumptions involved.

This investigation enhances our understanding of ULXs and UCXBs, offering a detailed view of the complex interplays within these enigmatic objects.  It also contributes to the broader understanding of GC dynamics, supported by simulations from the Monte Carlo cluster simulation, enhancing our comprehension of ULX and UCXB formation and evolution in dense stellar environments.

\section*{Acknowledgements}
We thank the referee for their helpful feedback. KCD acknowledges support for this work
provided by NASA through the NASA Hubble Fellowship grant HST-HF2-51528 awarded by the Space Telescope Science Institute, which is operated by the Association of Universities for Research in Astronomy, Inc., for NASA, under contract NAS5–26555. E.A. acknowledges support by NASA through the NASA
Hubble Fellowship grant HST-HF2-51501.001-A awarded by the Space Telescope Science Institute, which is operated by the Association of Universities for Research in Astronomy, Inc., for NASA, under contract NAS5-26555. 
SEZ acknowledges support from \textit{Chandra} grant GO3-24061X. AB acknowledges support for this project from the European Union's Horizon 2020 research and innovation program under grant agreement No 865932-ERC-SNeX. JS acknowledges support from NASA grant 80NSSC21K0628.
\section*{Data Availability}
The data underlying this article were accessed from \textit{Chandra} Data Archive (https://cda.harvard.edu/chaser/). The optical spectra can be found in the Gemini archive and the NOIRLab archive. 


\bibliographystyle{mnras}
\bibliography{ref} 




\bsp	
\label{lastpage}
\end{document}